\documentclass[]{spie}  
\usepackage{glossaries}
 
\usepackage{amsmath,amsfonts,amssymb}
\usepackage{graphicx}
\usepackage[colorlinks=true, allcolors=blue]{hyperref}

\usepackage{color,soul}

\usepackage[normalem]{ulem}
\usepackage{csvsimple}
\usepackage{booktabs}
\usepackage{colortbl}
\usepackage{xcolor}
\usepackage{xfrac}
\newcommand{\ra}[1]{\renewcommand{\arraystretch}{#1}}

\title{WFIRST Coronagraph Technology Requirements: Status Update and Systems Engineering Approach}

\author[a]{Ewan S. Douglas}
\author[a]{Ashley K. Carlton}
\author[a]{Kerri L. Cahoy}
\author[b]{N. Jeremy Kasdin}
\author[c]{Margaret Turnbull}
\author[d]{Bruce Macintosh}
\affil[a]{Space, Telecommunications, Astronomy, and Radiation Laboratory, Department of Aeronautics and Astronautics, Massachusetts Institute of Technology, 77 Massachusetts Avenue, Cambridge, MA, USA}
\affil[b]{Mechanical and Aerospace Engineering, Princeton University, Princeton, NJ, USA}
\affil[c]{SETI Institute, Carl Sagan Center for the Study of Life in the Universe, Off-Site: 2801 Shefford Drive, Madison, WI 53719, USA}
\affil[d]{Kavli Institute for Particle Astrophysics and Cosmology, Department of Physics, Stanford University, 382 Via Pueblo Mall, Stanford, CA USA}
\authorinfo{Further author information: (Send correspondence to E.S.D.)\\A.A.A.: E-mail: douglase@bu.edu}

 \newacronym{CGI}{CGI}{Coronagraph Instrument}
 \newacronym{IFS}{IFS}{Integral Field Spectrograph}
 \newacronym{HabEx}{HabEx}{Habitable Exoplanet Imaging Mission}
 \newacronym{LUVOIR}{LUVOIR}{Large Ultraviolet Visible and InfraRed Surveyor}
\begin{document} 
\maketitle
\begin{abstract}
The \gls{CGI} on the Wide-Field Infrared Survey Telescope (WFIRST) will demonstrate technologies and methods for high-contrast direct imaging and spectroscopy of exoplanet systems in reflected light, including polarimetry of circumstellar disks. 
The WFIRST management and \gls{CGI} engineering and science investigation teams have developed requirements for the instrument, motivated by the objectives and technology development needs of potential future flagship exoplanet characterization missions such as the NASA Habitable Exoplanet Imaging Mission (HabEx) and the Large UV/Optical/IR Surveyor (LUVOIR). 
The requirements have been refined to support recommendations from the WFIRST Independent External Technical/Management/Cost Review (WIETR) that the WFIRST CGI be classified as a technology demonstration instrument instead of a science instrument.

This paper provides a description of how the CGI requirements flow from the top of the overall WFIRST mission structure through the Level 2 requirements, where the focus here is on capturing the detailed context and rationales for the CGI Level 2 requirements. The WFIRST requirements flow starts with the top Program Level Requirements Appendix (PLRA), which contains both high-level mission objectives as well as the CGI-specific baseline technical and data requirements (BTR and BDR, respectively). Captured in the WFIRST Mission Requirements Document (MRD), the Level 2 CGI requirements flow from the PLRA objectives, BTRs, and BDRs. There are five CGI objectives in the WFIRST PLRA, which motivate the four baseline technical/data requirements. There are nine CGI level 2 (L2) requirements presented in this work, which have been developed and validated using predictions from increasingly refined observatory and instrument performance models. 

We also present the process and collaborative tools used in the L2 requirements development and management, including the collection and organization of science inputs, an open-source approach to managing the requirements database, and automating documentation. The tools created for the CGI L2 requirements have the potential to improve the design and planning of other projects, streamlining requirement management and maintenance. 

The WFIRST CGI passed its System Requirements Review (SRR) and System Design Review (SDR) in May 2018. The SRR examines the functional requirements and performance requirements defined for the system and the preliminary program or project plan and ensures that the requirements and the selected concept will satisfy the mission, and the SDR examines the proposed system architecture and design and the flow down to all functional elements of the system.

\end{abstract}

\glsresetall
\section{INTRODUCTION}
\label{sec:intro}

\subsection{WFIRST CGI Overview}

\begin{figure}
\centering
\includegraphics[width=8cm]{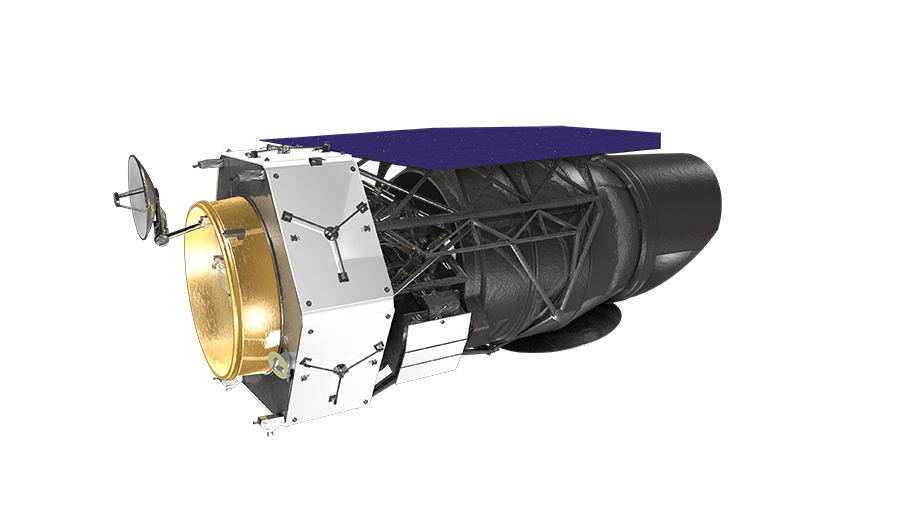}
\includegraphics[width=8cm]{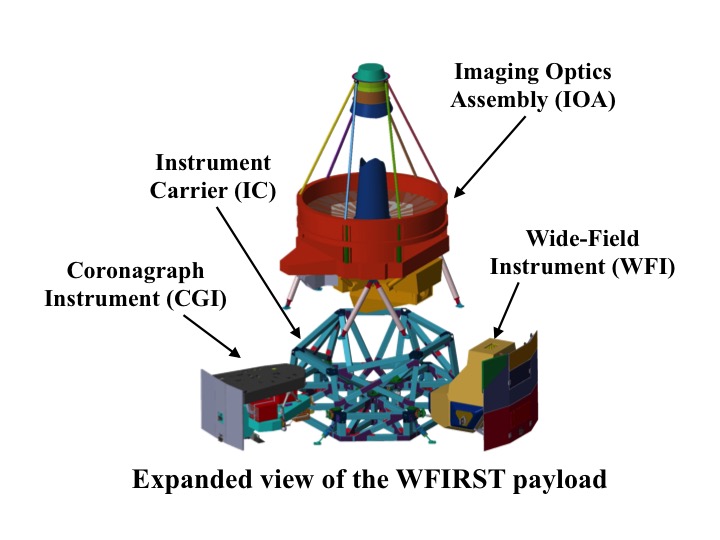}
\caption{Left: Artist's illustration of the WFIRST spacecraft. Right: Expanded view of the payload instruments on WFIRST. Image source: \url{https://wfirst.gsfc.nasa.gov/observatory.html}}\label{fig:wfirst_spacecraft}
\end{figure}

The Wide-Field Infrared Survey Telescope (WFIRST) is a NASA space-based observatory designed to address key questions in infrared astrophysics, dark energy science, and exoplanet detection \cite{Spergel2015_SDT}. The 2010 Decadal Survey\cite{Decadal2010} prioritized the WFIRST concept, and WFIRST is being developed for launch in the mid-2020's to orbit at the second Sun-Earth Lagrange point (SEL2). WFIRST has a 6-year planned mission duration, and Phase B of development began in April 2018. The WFIRST observatory has a 2.4-m diameter primary and two instruments (shown in Fig. \ref{fig:wfirst_spacecraft}. The primary science instrument is a wide-field infrared (WFI) instrument and the technology demonstration instrument is the coronagraph instrument,  \gls{CGI}. As the primary instrument, the WFI will observe billions of galaxies, measuring the history of cosmic acceleration and galactic evolution. The WFI will also perform a microlensing survey, probing the inner galaxy for thousands of exoplanets. 

This paper focuses on the WFIRST coronagraph instrument. \gls{CGI} will make high-contrast (of order 10$^{-9}$) direct imaging measurements and obtain moderate to low-resolution spectroscopy of exoplanets using a coronagraph to block light from the host star. 
This will also enable measurement of reflected brightness and polarization of circumstellar and protoplanetary disks. 
The CGI has two complementary coronagraphs: a Shaped Pupil Coronagraph (SPC) intended for exoplanet system and outer disk imaging\cite{Zimmerman2016} and a Hybrid Lyot Coronagraph (HLC) for exoplanet and inner disk imaging\cite{Trauger2016}. 
In addition to an imaging Electron Multiplying Charge-Coupled Device (EMCCD), the CGI has an \gls{IFS}\cite{rizzo_simulating_2017} which allows spectroscopic characterization of multiple targets simultaneously. An overview of the science possible with  CGI is discussed in Macintosh et al. (2017) and Kasdin et al. (2018)\cite{macintosh_science_2017,kasdin_wfirst_2018}.
The \gls{CGI} will demonstrate technology to enable future  missions: coronagraphy with active wavefront control; coronagraph elements, such a deformable mirrors\cite{sidick_effect_2015} and wavefront sensors\cite{shi_low_2016,shi_dynamic_2017}; wavefront sensing and control algorithms\cite{giveon_electric_2007,pueyo_optimal_2009,seo_hybrid_2017}; and high-contrast data processing algorithms\cite{ygouf_data_2015-1,ygouf_data_2016}.

\subsection{Requirements relationship to future missions} 
The CGI instrument status as a technology demonstration, rather than a science instrument, means requirements are drawn from the technological needs of science to be performed by future coronagraphic space telescopes. The WFIRST \gls{CGI} technology requirements are being developed to increase the technology readiness level and decrease risk for future flagship observatory missions such as the Habitable Exoplanet Imaging Mission \gls{HabEx} \cite{_habitable_,mennesson_habitable_2016} or the Large UV Optical Infrared Surveyor \gls{LUVOIR} \cite{_luvoir_,bolcar_initial_2016}. The WFIRST \gls{CGI} will demonstrate technologies such as wavefront control and understanding the effect of telescope stability in a space environment on high-contrast coronagraphic images. 

 \subsection{Requirements Definition Overview} 

Derived from the mission objective(s), requirements describe the necessary functions and features of the system. The requirements establish a basic agreement between the stakeholders and the developers. Requirements are useful for all stages of the mission lifecycle\cite{hirshorn_nasa_2017,SMAD}. Requirements are quantitative in nature and should not impose a solution, rather they establish a set of numerical requirements and constraints that correspond to the desired operational and functional capabilities that will meet the mission objective(s). Requirements follow a general hierarchy: mission goal/objective, top level and system-level requirements, and then subsystem requirements. 
Lower-level requirements are mapped to parent requirements, providing what is known as ``requirements traceability.'' 

\begin{figure}
\centering
\includegraphics[width=12cm]{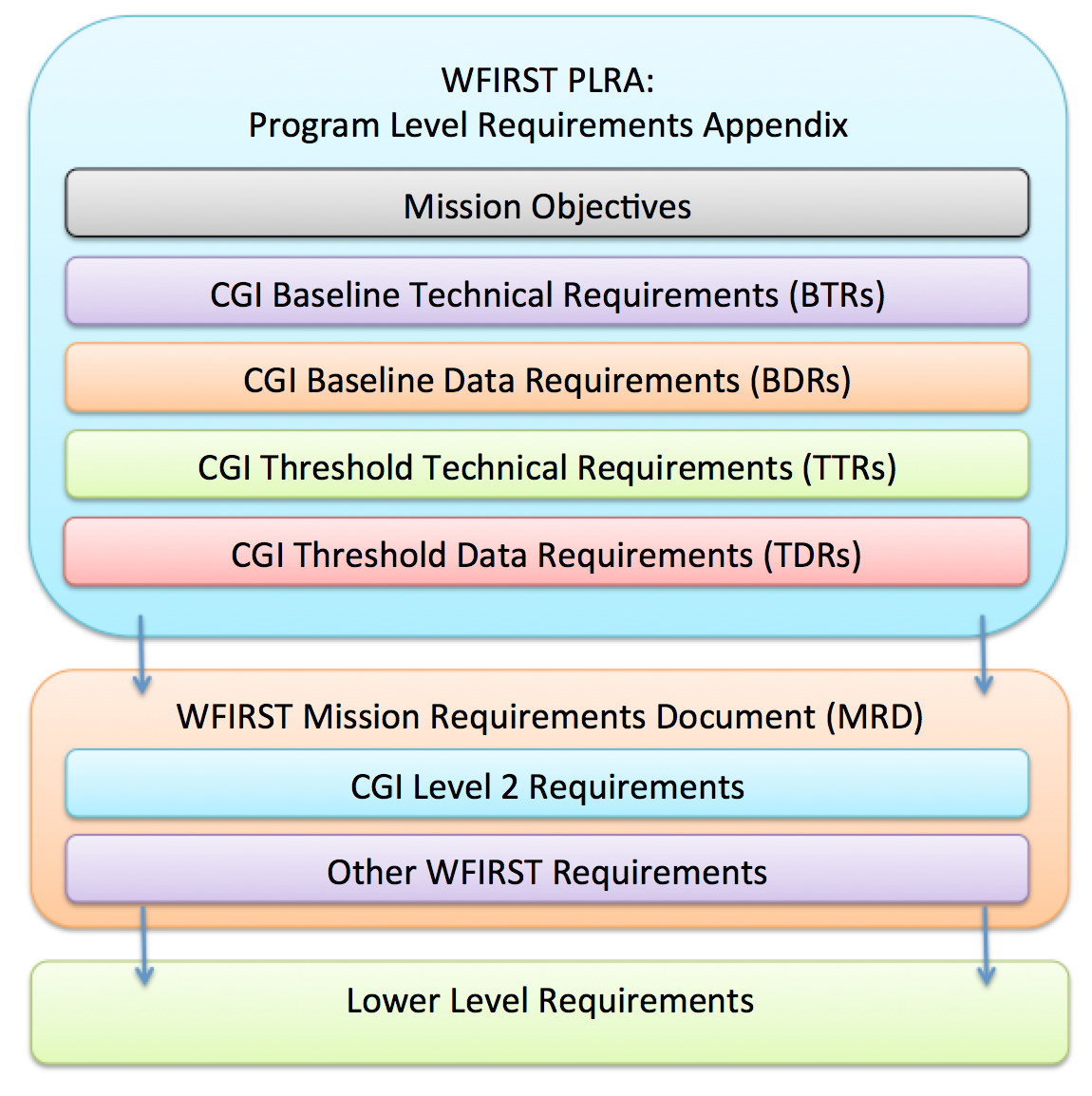}
\caption{High level requirements flow for the WFIRST mission. The mission objectives are described in the Program Level Requirements Appendix (PLRA). As a tech demo, the objectives relevant to the CGI are followed by a small number of baseline (and threshold) technical and data requirements. From there, the Level 2 requirements are developed as part of the WFIRST Mission Requirements Document (MRD). Lower-level requirements flow from the MRD down.}\label{fig:reqs_flow_CGI}
\end{figure}

There are three main types of requirements. Functional requirements are a statement of some\emph{thing} you need (a function). They can be such things as ``Payload data shall be communicated to the ground,'' or ``Desired orbit X shall be maintained for Y duration.'' There are also performance requirements, which define a \emph{characteristic} of something you need, \emph{i.e.,} how much of a function. The third type of requirement is constraints, which set limits on certain resources and define quantities that cannot be traded off with respect to cost, schedule, etc. Additional requirements (which generally fall within the previous three types) are interface, environmental, reliability, and safety requirements.

When writing requirements, one should aim for concise, explicit wording, often using ``shall'' statements and action verbs. 
All requirements must be quantifiable and verifiable. Each requirement must include the method by which the requirement will be verified, such as by test, by analysis, by inspection, and/or review of the design. There must be a rationale as to why that specific requirement exists. The rationale should also include any assumptions and document relationships (with links or report numbers) and any relevant design constraints. Requirements should also include revision dates and follow good engineering practices, such as including units. In summary, each requirement typically includes the following information: Requirement ID (for referencing purposes), level, requirement text, rationale, traced from / parent ID, owner/responsible party, verification method, and the date of last modification.
For more details on designing requirements, the reader is encouraged to review Chapter 4 of the \emph{NASA Systems Engineering Handbook}\cite{hirshorn_nasa_2017}.

\subsection{Systems Engineering challenges relevant to the WFIRST CGI Requirements}

Writing, maintaining, and verifying requirements can be difficult and time consuming. This section highlights a few of the key challenges as they are relevant to the \gls{CGI} requirements. Projects (and the scientists and engineers involved) are often under tight constraints due to resource limitations (e.g. cost). Program schedules, evolving objectives, and expanding scope often force analysis and modeling to happen concurrently with requirements development. While this iterative process of modifying requirements is to be expected, it can lead to out-of-date versions of requirements or, worse, can clobber previous requirements writing work and result in an inefficient use of time and resources. 

One challenge is not having clearly stated requirements, but rather vaguely stated requirements, sometimes driven by uncertainty. Examples of vague requirements include cases when the requirement is not verifiable, or it is unclear how the requirement supports a mission objective or another parent requirement. Requirements can also be over-specified, where they point to a particular solution instead of providing guidelines to the functionality required.

Another issue is when there is not clear responsibility for or ownership of requirements. For the CGI instrument, the Science Investigation Teams  has primary responsibility for the Level 2 requirements, and the Engineering Team has primary responsibility for the Level 3 requirements. Both teams contribute at each level, however, and include and incorporate feedback from each other in the requirements development process. 

The most persistent issue during the lifecycle of a program is  ``requirements creep,'' a term used to describe the increase in number, complexity, or scope of requirements over time. Any out of sequence or delayed change to requirements for a program negatively affects cost and schedule. 

\subsection{Requirements Management Tools}

A variety of tools have been employed to keep requirements up-to-date and minimize creep. These including Microsoft Excel, Dynamic Object Oriented Requirements System (DOORS), rmToo, Model-Based System Engineering (MBSE), and others.
Each approach has strengths and weaknesses and some are briefly described below. 

A  spreadsheet (or collection of sheets) can be used as a relational database for requirements, with the benefit that it is easy to parse and the software, such as Microsoft Excel and OpenOffice, is familiar to many people.
 However, spreadsheets are challenging due to version control (passing spreadsheets around for editing), the number of sheets needed for larger projects, and, on its own, spreadsheet software does not automatically produce a corresponding formal requirements document.

The Dynamic Object Oriented Requirements Management System (DOORS) is an commercial object-oriented database, where each requirement is an ``object,'' rather than a row in a relational table \cite{DOORS}. 
DOORS is widely used in aerospace engineering. However, the system requires purchased licensing and this may limit the number of active participants in the requirements development and editing process.

Model-Based Systems Engineering (MBSE) tools for requirements management integrate models, analyses, budgets, and other system constraints together. MBSE creates a fully traceable system of requirements and interactions and is able to capture dependencies. In using MBSE tools, there can be logistical and technical challenges, such as organizational and cultural change and training.   \cite{Bayer_2012,Bayer_2013}. 

rmToo is an open-source requirements management tool\footnote{http://rmtoo.florath.net/}. rmToo is a command line tool, relying only on input and output files rather than a special tool set environment. rmToo also allows for  version control by integrating with git. 

The WFIRST CGI requirements development  started with a spreadsheet based approach for sharing  text between the instrument  and the science investigation teams.
These spreadsheets became  unwieldy as the requirements process evolved, more stakeholders were included, and the number of requirements and ancillary data increased.
To manage the growing body of requirement and enable more rapid revisions and collaborations, we adopted a set of opensource tools built on a decentralized version control system and scripted document generation, which will be detailed in Sec. \ref{sec:approach}.
 
\section{CGI requirements}
The  WFIRST requirements flow starts with the top Program Level Requirements Appendix (PLRA), which contains both high-level mission objectives as well as the CGI-specific baseline technical and data requirements (BTR and BDR, respectively). 
Captured in the WFIRST Mission Requirements Document (MRD), the Level 2 CGI requirements flow from the PLRA objectives, BTRs, and BDRs. There are five CGI objectives in the WFIRST PLRA, which motivate the four baseline technical/data requirements. 
The text of the CGI level 2 (L2) requirements are summarized in Table \ref{tab:L2_reqs}, which have been developed and validated using predictions from increasingly refined observatory and instrument performance models in order to best satisfy the mission technology demonstration goals. 
The current requirements, along with detailed rationales and verification plans, are also listed in Appendix \ref{app:L2}.

\begin{figure}
\centering
\includegraphics[width=16cm]{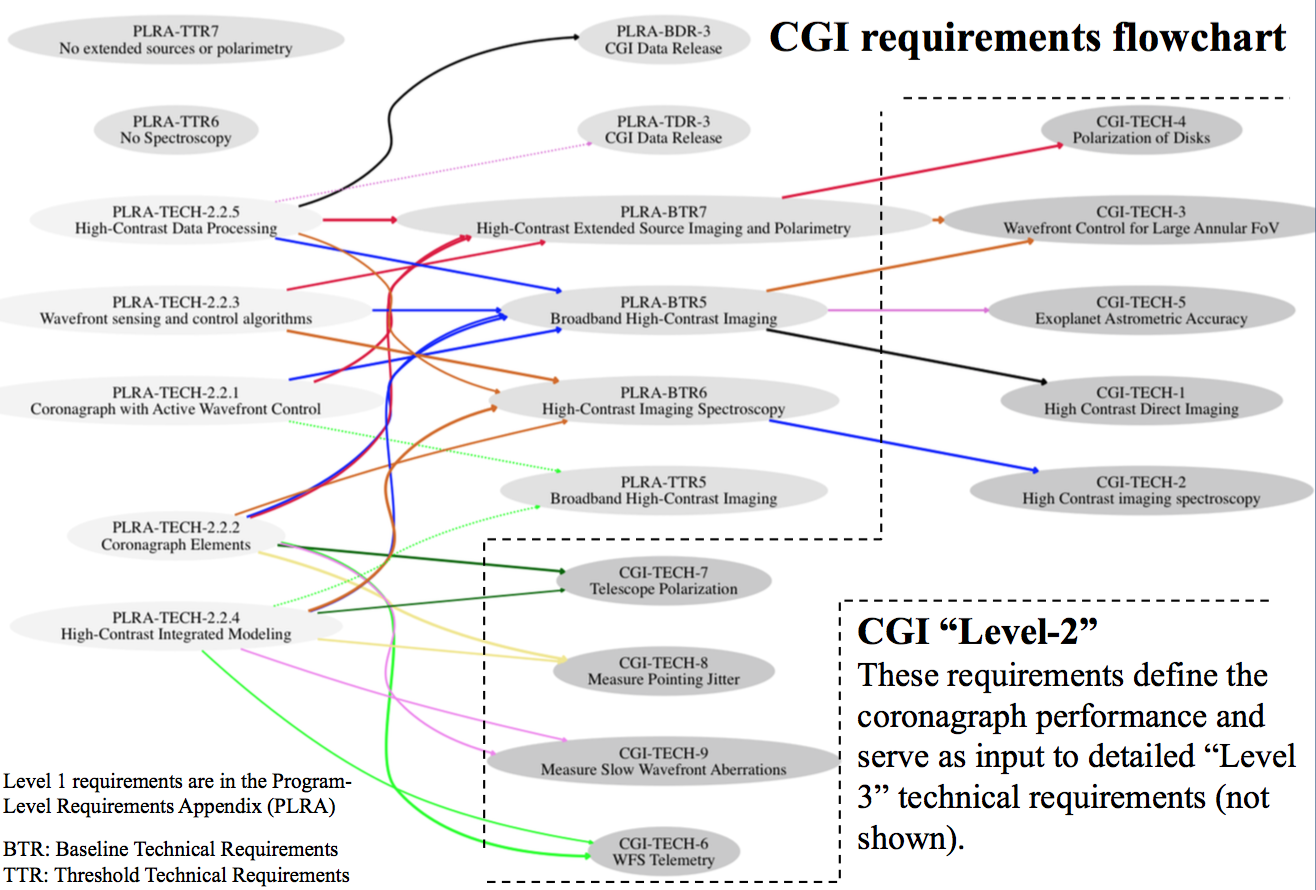}
\caption{WFIRST CGI requirements flow from objectives to Level 2. Level 2 requirements are outlined by the dashed line.  Version 4.8, \textit{git commit}:1a54825. This flowchart was automatically generated using Graphviz and Doorstop (see Section \ref{sec:approach}).}\label{fig:cgiflowchart}
\end{figure}

\begin{table*}\centering
\caption{Summary of WFIRST CGI Level 2 Requirements, v4.8, \textit{git commit}:1a54825}\label{tab:L2_reqs}
\ra{1.3}
\begin{tabular}{p{4.5cm} p{11cm}}\toprule
Requirement Title & WFIRST CGI shall be able to:  \\
\midrule
\rowcolor{black!10} High Contrast Direct Imaging (CGI-TECH-1) & ...measure the brightness of an astrophysical point source to an SNR of 10 or greater within 10 hours of integration time on the target in CGI Filter Band 1 for an object with a source-to-star flux ratio as faint as 1e-7 at separations from 0.16'' to 0.21'', 5e-8 at separations from 0.21'' to 0.4'', and 1e-7 at separations from 0.4'' to 0.45'' \\
High Contrast Imaging Spectroscopy (CGI-TECH-2) & ...measure spectra of an astrophysical point source with R = 50 or greater spectral resolution with a wavelength accuracy of 2 nm or smaller to an SNR of 10 within 100 hours of integration time on the target in CGI Filter Band 3 for an object with a source-to-star flux ratio as faint as 1e-7 at separations from 0.21'' to 0.27'', 5e-8 at separations from 0.27'' to 0.53'', and 1e-7 at separations from 0.53'' to 0.60''. \\
\rowcolor{black!10} Wavefront Control for Large Annular FoV (CGI-TECH-3)  & ...measure the brightness around a star as faint as V = 5 mag with an SNR of 10 or greater within 24 hours of integration time on the target in CGI Filter Band 4 for an extended source with an integrated surface brightness per resolution element equivalent to a source-to-star flux ratio as faint as 1e-7 at separations from 0.47'' to 0.54'', 5e-8 at separations from 0.54'' to 1.36'', and 1e-7 at separations from 1.36'' to 1.44''.
 \\
Polarization of Disks (CGI-TECH-4)  & ...map the linear polarization of a circumstellar debris disk that has a polarization fraction greater or equal to 0.3 with an uncertainty of less than 0.03 in CGI Filter Band 1 and CGI Filter Band 4, assuming an SNR of 100 per resolution element.
 \\
\rowcolor{black!10} Exoplanet Astrometric Accuracy (CGI-TECH-5)  & ...measure the relative astrometry between an astrophysical point source and its host star, in photometric images, for separations from 0.21'' to 1.36'', with an accuracy of 5 milliarseconds or less, assuming an SNR of 10 or greater, including systematic errors.
 \\
WFS Telemetry (CGI-TECH-6) & ...capture wavefront control system telemetry concurrently with science data, including raw wavefront sensor measurements and commanded deformable mirror actuator values. \\
\rowcolor{black!10} Telescope Polarization (CGI-TECH-7)  & ...measure the complex electric fields of incident light in two orthogonal polarization states. \\
Measure Pointing Jitter (CGI-TECH-8) & ...measure observatory tip/tilt disturbances at the CGI occulter at frequencies from 0.1 Hz to 100 Hz with accuracy better than 0.5 mas rms on sky per axis for a V=2 mag or brighter star. \\
\rowcolor{black!10} Measure Slow Wavefront Aberrations (CGI-TECH-9)  & ...estimate the average rate of change over 1 hour period at the CGI occulter for each of focus, astigmatism, coma, trefoil, and 3rd-order spherical aberrations, with accuracy better than 0.1 nm/hour, when pointed at a V=2 mag or brighter star. \\
\bottomrule
\end{tabular}
\end{table*}

\subsection{Level 3 requirements}

The next more detailed set of requirements (Level 3) is under development by the instrument team\cite{poberezhskiy_wfirst_2018} and will more precisely define  the instrument parameters needed to achieve the measurements detailed above.

\section{APPROACH}\label{sec:approach}
\subsection{Science Yield Tools}
In the process of exploring the \gls{CGI} technical requirements, a variety of science tools for characterizing the science impact of design changes were developed by the science investigation teams. These tools quantify the number of potential targets and provide information on how to optimize the system to operate in regimes with maximum science yield for future missions.

\subsubsection*{Yield Models}
Several models to quantify the expected exposure time to detect radial velocity detected exoplanets and yield of previously undiscovered exoplanets have been developed. Nemati et al.\cite{nemati_sensitivity_2017} developed an analytic model for the estimation of exposure times for exoplanet imaging and spectroscopy in the presence of speckles.
Savransky et al.\cite{savransky_exosims_2017,savransky_wfirst-afta_2016} developed an open-source mission simulation tool for which incorporates occurrence rates and mission yield optimization routines.
Both of these models use raw instrument contrast and throughput curves calculated using an end-to-end numerical model of the instrument, including dynamic telescope wavefront error\cite{krist_overview_2015,krist_wfirst_2017}.

\subsubsection*{Debris Disk Sensitivity}
Models of coronagraph performance were adapted to extended sources to approximate the sensitivity to scattered light from circumstellar debris disks \cite{schneider_quick_2014,debes_wfirst_2018}.

\subsubsection*{Spectroscopic Retrieval}
The WFIRST \gls{IFS} \cite{gong_flight_2017} provides imaging at moderate spectral resolution for spectroscopic characterization and wavelength dependent wavefront control. To explore the relationship between measurable giant planet properties and requirements on the integral field spectrometer, an atmospheric retrieval method was developed and run for different realizations of \gls{IFS} spectral resolution and signal-to-noise ratios for synthetic as well as Jupiter and Saturn analog exoplanets\cite{marley_quick_2014,lupu_developing_2016,hildebrandt_wfirst_2018}.
\subsubsection*{Orbital Modeling}
To assess the astrometric precision required, orbit fitting with Bayesian rejection sampling\cite{blunt_orbits_2017} was used to establish the astrometric precision and signal-to-noise ratios required to recover directly imaged exoplanet orbital parameters.

\subsection{Requirement Management Tools}

Initial WFIRST CGI requirements management was conducted using spreadsheets.
While straightforward and widely supported, managing requirements solely in spreadsheet form led to several challenges, particularly version control, distributions, and dependency visualization.

Managing versions across a large, diverse team becomes problematic when multiple contributors have concurrently modified requirements. 
A modern decentralized version control system such as \textit{git}\cite{_git_2018,Blischak_quick_2016} or \textit{mercurial}\cite{_mercurial_}, where revisions are tracked  with automatic character-by-character change logs, greatly simplifies merging divergent requirements documents. 
Dependency visualization, creating a flowchart of interdependencies and links between requirements, allows intuitive and rapid vetting of requirement structure and link relevance.
Distribution of the requirements in spreadsheet form is inconvenient for presentation and sharing with stakeholders. For verbal presentations, slides are typically more useful, while manuscript form documents are often needed for formal record keeping. 

To overcome these challenges, we built an infrastructure around the Doorstop\cite{browning_doorstop_2014} Python requirements management tool. 
Doorstop imports a requirements management spreadsheet and parses it into a tree of human-readable YAML files (one per requirement), tracking and validating links between requirements, and publishing hyperlinked requirements documents in  HTML or Markdown formats.
In addition to these features, we developed a suite of scripts to leverage Doorstop's support for link tracking and version control. 
The doorstop API allows easy parsing of links to autogenerate dependency visualization using Graphviz\cite{ellson_graphviz_2001,bank_graphviz_2018,_graphviz_}.
The Graphviz \textit{dot} tool positions nodes to minimize the number of edge (connecting lines) crossings and edge length \cite{ellson_graphviz_2001}, producing
\textit{dot}  graphs to visualize the requirements flow as illustrated for WFIRST in  Fig. \ref{fig:cgiflowchart}.

\begin{figure}[t]
\centering
\includegraphics[width=12cm]{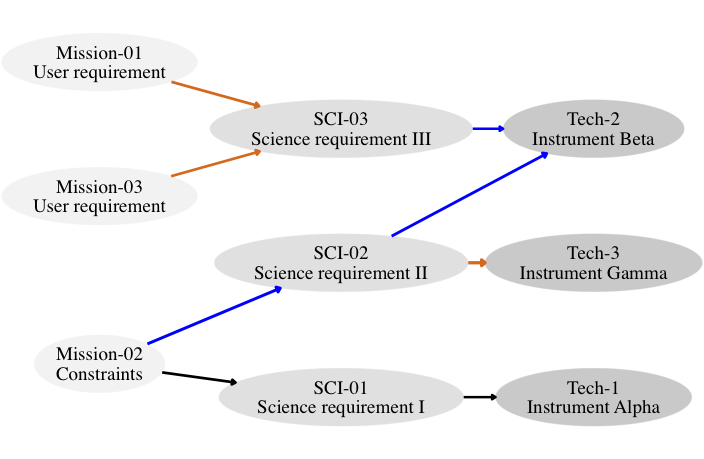}
\caption{Illustrative example of generic requirements generated by Doorstop using Graphviz.}\label{fig:reqs_flow_Graphviz}
\end{figure}

An example requirements  flow is shown in Fig. \ref{fig:reqs_flow_Graphviz}, representing a simple three level flow from mission to science and technical requirements. 
An input table from one level of the example requirements is shown in Table \ref{tab:example_l2}.
Higher level requirements start on the left with light gray and lower level requirements are on the far right in darker gray.
Different line-weights and colors connect to each downstream requirements, helping to trace the upstream parents of a requirement in complex requirements documents.
A template to reproduce this figure using the tools described here is available via Github\footnote{\url{https://github.com/douglase/doorstop_requirements_template}} and archived using Zenodo\cite{douglase_douglase/doorstop_requirements_template_2017}. 
Other features include customized markdown output files using Pandoc\cite{macfarlane_pandoc_2017} which allows  of simple hyperlinked publication markdown pages \footnote{e.g. for realtime publication on \url{https://www.github.com}} as well as automated generation of presentation slides using LaTeX Beamer format,  annotated with a unique revision number and git commit hash for traceability.

\begin{table}
\begin{tiny}
\begin{filecontents*}{sci_L2.csv}
uid,level,text,links,short name,rationale,verification plan
SCI-01,1,Mission shall be able to measure x to y precision,Mission-02,Science requirement I,,
SCI-02,1,Mission shall be able to measure xy to yz precision,Mission-02,Science requirement II,,
SCI-03,1,Mission shall be able to measure xz to yz precision,"Mission-01\,Mission-3",Science requirement III,,
\end{filecontents*}
\csvautotabular{sci_L2.csv}\label{tab:example_l2}
\end{tiny}
\caption{Illustrative example of a generic Level 2 requirements table in progress with empty columns for rationale and verifications plan.}
\end{table}

\section{Conclusions}
The WFIRST CGI Level 2 requirements capture the measurements required to demonstrate high-contrast imaging of exoplanets in reflected light with a space telescope. 
We have presented a systematic approach for requirements development, along with a flexible, version controlled tool that  automates and simplifies visualization requirements continuously through the development process. 
This combination has streamlined  requirements development, eased tracking and merging conflicting versions, allowing the WFIRST \gls{CGI} requirements development team to focus on the structure and content of the requirements.  

As the mission is refined, the presented requirements are expected to evolve and serve as a groundwork for future instrument development during the ongoing Phase B of the mission. 





\acknowledgments 
The authors acknowledge valuable inputs from Ilya Poberezhskiy,  Vanessa Bailey, and the rest of the JPL CGI team.
 Support for this work was provided by the WFIRST Science Investigation team prime award \#NNG16PJ24C.
 AKC was also supported by a NASA Space Technology Research Fellowship \#NNX16AM74H.

\appendix
\section{WFIRST CGI Level 2 CGI Requirements}\label{app:L2}
The current (last edited April 19, 2018) CGI Level 2 requirements, rationales, links to higher level requirements, and verification plans, with the unique git commit identifier \textit{1a54825} and tagged as v4.8. References for citations are included in the bibliography of this manuscript \cite{noll_zernike_1976,vogt_coronagraphic_2011,stark_lower_2015,singh_psf_2016}.

\begin{small}
\subsection*{High Contrast Direct Imaging CGI-TECH-1}

WFIRST CGI shall be able to measure the brightness of an astrophysical
point source to an SNR of 10 or greater within 10 hours of integration
time on the target in CGI Filter Band 1 for an object with a
source-to-star flux ratio as faint as 1e-7 at separations from 0.16
arcsec to 0.21 arcsec, 5e-8 at separations from 0.21 arcsec to 0.4
arcsec, and 1e-7 at separations from 0.4 arcsec to 0.45 arcsec.

\emph{Rationale:}\\
This requirement demonstrates coronagraphy at small working angles at a
contrast level that will require wavefront control. The use of three
separation regions relax requirements at the inner and outermost regions
which are consistent with masks that meet the technology demonstration
objectives, such as an HLC mask, where tip/tilt and low-order errors can
affect performance close to the inner and outer working angles. We
assume a point source to be any astrophysical object whose angular size
is such that it cannot be resolved by WFIRST CGI, with a background flux
less than or equal to the solar zodiacal dust, for a star as faint as V
= 5 mag with a stellar radius of less than or equal to 0.4
milliarcseconds. We use ``point source'' here, since a background object
and a companion exoplanet could both be observed. Additional
information, such as multiple observations, may be required to
distinguish the two. The filter band referenced is defined in the WFIRST
CGI Filter Table which is controlled as part of the WFIRST Mission
Requirements Document (TBD document \#). A 23 V-mag per square arcsecond
solar zodiacal background is assumed (Stark et al. 2015). We assume a
gain of no greater than 2 from post-processing.

\emph{Verification plan:}\\
This requirement will be verified by test and by analysis, using
hardware testbeds and simulators.

\emph{Parent links:}{PLRA-BTR5}

\subsection*{High Contrast imaging spectroscopy CGI-TECH-2}

WFIRST CGI shall be able to measure spectra of an astrophysical point
source with R = 50 or greater spectral resolution with a wavelength
accuracy of 2 nm or smaller to an SNR of 10 within 100 hours of
integration time on the target in CGI Filter Band 3 for an object with a
source-to-star flux ratio as faint as 1e-7 at separations from 0.21
arcsec to 0.27 arscec, 5e-8 at separations from 0.27 arcsec to 0.53
arcsec, and 1e-7 at separations from 0.53 arcsec to 0.60 arcsec.

\emph{Rationale:}\\
This requirement demonstrates coronagraphic spectroscopy at small
working angles and at a contrast level that will likely require
wavefront control. The use of three separation regions relax
requirements at the inner and outermost regions which are consistent
with masks that can meet the technology demonstration objectives, such
as an SPC mask, where tip/tilt and low-order errors can affect
performance close to the inner and outer working angles. We assume a
point source to be any astrophysical object whose angular size is such
that it cannot be resolved by WFIRST CGI, a scattered light background
equal to the solar zodiacal dust, for star as faint as V = 5 mag with a
stellar radius of less than or equal to 0.4 milliarcseconds. We use
``point source'' here, since a background object and a companion
exoplanet could both be observed. Additional information, such as
multiple observations, may be required to distinguish the two. The
filter band referenced is defined in the WFIRST CGI Filter Table which
is controlled as part of the WFIRST Mission Requirements Document (TBD
document \#). We assume a gain of no greater than 2 from
post-processing.

\emph{Verification plan:}\\
This requirement will be verified by test and by analysis, using
hardware testbeds and simulators.

\emph{Parent links:} {PLRA-BTR6}

\subsection*{Wavefront Control for Large Annular FoV CGI-TECH-3}

WFIRST CGI shall be able to measure the brightness around a star as
faint as V = 5 mag with an SNR of 10 or greater within 24 hours of
integration time on the target in CGI Filter Band 4 for an extended
source with an integrated surface brightness per resolution element
equivalent to a source-to-star flux ratio as faint as 1e-7 at
separations from 0.47 arcsec to 0.54 arcsec, 5e-8 at separations from
0.54 arcsec to 1.36 arcsec, and 1e-7 at separations from 1.36 arcsec to
1.44 arcsec.

\emph{Rationale:}\\
This requirement demonstrates coronagraphy at large working angles,
which will drive the number of actuators on the DM, and coronagraphic
sensitivity to image disks and exozodiacal dust. The performance in this
requirement is assumed to be achieved after postprocessing, and
postprocessing gains will be different for point-like and extended
sources. Spatial resolution is defined as Nyquist sampling the
diffraction limit of the Optical Telescope Assembly (OTA) at the
shortest wavelength in the CGI Filter Table. The sensitivities per
resolution element are chosen to be consistent with masks that can meet
the technology demonstration objectives, such as a SPC disk mask. The
use of three separation regions relax requirements at the inner and
outermost regions, where tip/tilt and low-order errors can affect
performance close to the inner and outer working angles. The filter band
referenced is defined in the WFIRST CGI Filter Table which is controlled
as part of the WFIRST Mission Requirements Document (MRD-450). We assume
a gain of no greater than 2 from post-processing. Resolution element is
defined as the solid angle within the half-max contour of the
coronagraph PSF.

\emph{Verification plan:}\\
This requirement will be verified by test and by analysis, using
hardware testbeds and simulators.

\emph{Parent links:} {PLRA-BTR5},
{PLRA-BTR7}

\subsection*{Polarization of Disks CGI-TECH-4}

WFIRST CGI shall be able to map the linear polarization of a
circumstellar debris disk that has a polarization fraction greater or
equal to 0.3 with an uncertainty of less than 0.03 in CGI Filter Band 1
and CGI Filter Band 4, assuming an SNR of 100 per resolution element.

\emph{Rationale:}\\
This requirement demonstrates CGI can achieve calibration accuracy
needed to recover dust properties and supports polarization-differential
imaging (PDI).~ As angular differential imaging is often unsuitable for
disk postprocessing due to self-subtraction, PDI is expected to provide
the expected postprocessing gains for extended source requirements.~ CGI
Filter Band 1 is expected to be used with the HLC disk mask, and CGI
Filter Band 4 is expected to be used the CGI SPC disk mask.

\emph{Verification plan:}\\
This requirement will be verified by test and by analysis, using
hardware testbeds and simulators.

\emph{Parent links:} {PLRA-BTR7}

\subsection*{Exoplanet Astrometric Accuracy CGI-TECH-5}

WFIRST CGI shall be able to measure the relative astrometry between an
astrophysical point source and its host star, in photometric images, for
separations from 0.21 arcsec to 1.36 arcsec, with an accuracy of 5
milliarcsec or less, assuming an SNR of 10 or greater, including
systematic errors.

\emph{Rationale:}\\
This requirement demonstrates measurement of orbital parameters to
inform multiple epoch imaging/followup. This requirement assumes that it
will be possible to obtain knowledge of the source-to-star position
angle from concurrent measurements by other WFIRST sensors (e.g.
detectors, ADCS sensors). Rejection sampling orbital retrieval modeling
shows that 5 mas accuracy at SNR=10 is sufficient to recover the orbital
parameters required, assuming astrometric error is inversely
proportional to the SNR of a point source. The SNR is assumed to be
\textasciitilde{}10 at the brightest point in the orbit and is degraded
over the orbit following a Lambert phase function using the methods in
Blunt et al. (2017). The separation of 0.21 arcseconds is 1 lambda/D
outside of the IWA in CGI Filter Band 1 due to the JPL request that
precise astrometric centroiding not be required at the IWA.

\emph{Verification plan:}\\
Verify by analysis, by injection and recovery of simulated planets into
CGI camera data after instrument integration.

\emph{Parent links:} {PLRA-BTR5}

\subsection*{WFS Telemetry CGI-TECH-6}

WFIRST CGI shall be able to capture wavefront control system telemetry
concurrently with science data, including raw wavefront sensor
measurements and commanded deformable mirror actuator values.

\emph{Rationale:}\\
This requirement demonstrates that CGI telemetry will meet its mission
objective to inform operators of the current operational status of the
instrument as well as provide additional information to support
postprocessing and modeling validation and refinement. Laboratory work
done on other coronagraphic instruments (e.g. Vogt et al. 2011, Singh et
al. 2016) have shown that wavefront sensing telemetry can be used post
facto to improve coronagraph PSF subtraction. A detailed list of
telemetry and necessary rates and parameters is captured in WFIRST CGI
Engineering Telemetry Document (TBD Document \#), and includes: all
camera frames, readings, settings, timestamps; tachometer data for all
reaction wheels; thermal sensor readings and heater settings; mechanism
commands and encoder readings with timestamps; power status of all
powered items; deformable mirror commands and actuator voltages; all CGI
related command responses/faults, intermediate engineering products
generated by CGI, commands sent by CGI pointing control to observatory
ACS; fault or anomaly status of CGI.

\emph{Verification plan:}\\
This requirement will be verified by test, using simulated telemetry
streams as necessary.

\emph{Parent links:} {PLRA-TECH-2.2.2},
{PLRA-TECH-2.2.4}

\subsection*{Telescope Polarization CGI-TECH-7}

WFIRST CGI shall be able to measure the complex electric fields of
incident light in two orthogonal polarization states.

\emph{Rationale:}\\
We need to measure the effects of polarization on contrast by measuring
complex electric fields in two orthogonal states. Measuring ``contrast''
in multiple polarization states alone is not sufficient because the
contrast in orthogonal states can be the same, but the underlying
electric field will have opposite sign. For example, measuring contrast
alone, in two orthogonal pol states, doesn't by itself discriminate
other sources of incoherent light. Polarization-dependent aberrations
from mirror coatings have not been previously measured at the levels
necessary for exoplanet direct imaging. While future coronagraphic
missions will likely not have the same polarization-aberration
environment as WFIRST due to slower optics, WFIRST CGI will improve the
modeling uncertainties on polarization aberration terms for their error
budgets.

\emph{Verification plan:}\\
This requirement will be verified by test and analysis: measurement on
the flight~CGI~with a telescope simulator GSE. Note: It will not be
feasible to verify Tech-7 on the ground with the real OTA (partial ACF
coverage, insufficient facility stability)

\emph{Parent links:} {PLRA-TECH-2.2.2},
{PLRA-TECH-2.2.4}

\subsection*{Measure Pointing Jitter CGI-TECH-8}

WFIRST CGI shall be able to measure observatory tip/tilt disturbances at
the CGI occulter at frequencies from 0.1 Hz to 100 Hz with accuracy
better than 0.5 mas rms on sky per axis for a V=2 mag or brighter star.

\emph{Rationale:}\\
We will measure pointing disturbances as they are presented at the
coronagraph mask over a range of timescales for fairly bright targets so
that there are enough photons for a high SNR (high quality) measurement.
This requirement demonstrates that CGI will meet its mission objective
to characterize jitter disturbances of the WFIRST OTA.

\emph{Verification plan:}\\
By test and analysis. During CGI integration and test, measurements will
be made with jitter injection stable to \textless{} 0.1 Hz over 100
seconds from a calibrated upstream mirror in the telescope simulator.
Tip/tilt disturbance measurements will be made for at least one
coronagraph mask with the CGI LOWFS sensor.

\emph{Parent links:} {PLRA-TECH-2.2.2},
{PLRA-TECH-2.2.4}

\subsection*{Measure Slow Wavefront Aberrations CGI-TECH-9}

WFIRST CGI shall be able to estimate the average rate of change over 1
hour period at the CGI occulter for each of focus, astigmatism, coma,
trefoil, and 3rd-order spherical aberrations, with accuracy better than
0.1 nm/hour, when pointed at a V=2 mag or brighter star.

\emph{Rationale:}\\
This requirement responds to PLRA BTR8 (Engineering Data Collection and
Performance Characterization).~ Measurement of wavefront drift at these
frequencies and magnitudes will be relevant to future coronagraphic
space observatories, as the required magnitudes for coronagraphic
performance are comparable; for example the HabEx VVC8 concept requires
all of focus, astigmatism, coma, trefoil, and spherical to be less than
0.2 nm.~ The wavefront error terms listed should be taken as equivalent
to Noll Zernikes Z4-Z11, as defined in (Noll, 1976).~ Unlike tip/tilt,
low-order wavefront drifts are not expected to be cyclic, and measuring
linear trends is more relevant for future observatory planning. The
average rate of change of wavefront aberrations over a period on the
order of 1 hour is what we care about in the RDI and ADI observation
context. This term most directly feeds into CGI error budgets.

\emph{Verification plan:}\\
This requirement will be verified by test and analysis. For example,
injection of low-order drift from rigid body motions of optics within
the telescope simulator.

\emph{Parent links:} {PLRA-TECH-2.2.2},
{PLRA-TECH-2.2.4}

\end{small}

\bibliographystyle{spiebib}
\bibliography{report,wfirst} 

\end{document}